\documentclass[prl,aps,twocolumn,showpacs,tightenlines,amssymb]{revtex4}

\usepackage{epsfig}

\begin{document}

\title{Hybrid noiseless subsystems for quantum communication over optical
fibers}

\author{Jonathan L. Ball}
\author{Konrad Banaszek}
\affiliation{Centre for Quantum Computation, Clarendon Laboratory,
University of Oxford, Oxford OX1 3PU, United Kingdom}

\date{\today}

\begin{abstract}
We derive the general structure of noiseless subsystems for optical
radiation contained in a sequence of pulses undergoing collective
depolarization in an optical fiber. This result is used to identify
optimal ways to implement quantum communication over a collectively
depolarizing channel, which in general combine various degrees of
freedom, such as polarization and phase, into joint hybrid schemes
for protecting quantum coherence.
\end{abstract}


\maketitle

The ability to maintain and transmit coherent quantum superpositions
is a prerequisite for the realization of protocols in quantum
information processing. In communication tasks, such as quantum key
distribution \cite{GisiRiboRMP02} and teleportation
\cite{BouwPanNAT97}, a natural choice for the physical carrier of
quantum information is optical radiation, often confined to optical
fibers in order to minimize uncontrolled interactions with the
environment. Quantum states encoded in low-intensity, narrowband
optical fields experience two dominant decoherence mechanisms when
transmitted through optical fibers \cite{GisiRiboRMP02}. The first
mechanism is a linear loss of the light amplitude through residual
absorption, scattering and coupling out to non-guided modes. The
second one is a random transformation of the polarization state of
the transmitted light due to changes in fiber birefringence caused
by environmental conditions such as temperature fluctuations and
aeolian vibrations. In this paper we address the latter mechanism
and analyze ways to protect quantum coherence against depolarization
by using general, multiphoton states of optical radiation. We assume
here that linear losses are polarization-independent and that they
can be dealt with in a standard way by postselecting transmissions
when no photons have been lost.

What makes fiber depolarization manageable in quantum communication
tasks is that its temporal variations are relatively slow. In
practice, the fiber birefringence remains virtually constant for
long sequences of pulses that can be temporally resolved by the
receiver, or even for roundtrip transmissions from the receiver to
the sender and back. This enables a number of strategies for
implementing communication protocols over such collectively
depolarizing channels. The first one is to use a degree of freedom
that is not affected by depolarization at all, such as the relative
phase between consecutive pulses. This idea is best illustrated by
time-bin encoding used successfully in a variety of experiments
\cite{TimeBin}. A practical issue in phase-encoding is the necessity
to maintain a suitable phase reference, which can be accomplished by
active stabilization of distributed interferometers
\cite{HughMorgJMO00}, two-way roundtrip communication
\cite{PlugandPlay}, or one-way autocompensation using photon pairs
\cite{OWA}. The second recently proposed strategy exploits the
advances in the theory of noiseless subsystems \cite{Noiseless} to
construct joint polarization states of several photons that are
immune to collective depolarization
\cite{BoilGottPRL04,CabePRL03,BourEiblPRL04}. From the fundamental
perspective, such states enable quantum communication without a
shared reference frame \cite{BartRudoPRL03}. Most of the work in
this direction is based on results obtained for an ensemble of
qubits that undergo an identical symmetrized interaction with the
environment. In the optical domain, this model translates into a
sequence of pulses containing one photon each and prepared in joint,
usually entangled, polarization states.

Sequences of identically polarized pulses employed in phase
encoding, or trains of single-photon pulses used to implement
noiseless subsystems, form very particular classes of states in the
entire Hilbert space describing a set of optical modes travelling in
a fiber. The central aim of this paper is to analyze the structure
of the complete Hilbert space under collective depolarization and to
determine its potential for implementing quantum communication
protocols. This approach will provide a general treatment of
depolarization in optical fibers that includes a number of
previously proposed schemes as special cases. Furthermore, our
analysis will reveal the existence of hybrid noiseless subsystems
that combine both polarization and multiphoton phase encoding in a
non-trivial way. These hybrid subsystems are optimal in terms of
required resources, namely the total numbers of temporal slots and
photons.

We shall consider general, multiphoton states of optical radiation
prepared over $N$ temporal slots that experience identical
birefringence. The polarization transformation between a pair of
annihilation operators describing two orthogonally polarized modes
that occupy one temporal slot is given by an element ${\bf\Omega}$
of the Lie group U(2). The element ${\bf\Omega}$ can be decomposed
into a product of the overall phase factor, which we will denote by
$e^{-i\alpha({\bf\Omega})}$, and the remaining matrix
$e^{i\alpha({\bf\Omega})}{\bf\Omega}$ with the determinant equal to
one which belongs to SU(2). The element ${\bf \Omega}$ induces a
unitary transformation $\hat{U}({\bf \Omega})$ in the corresponding
two-mode Fock space. For our purposes it is natural to use the
Schwinger representation \cite{SchwingerRepresentation} in which the
Hilbert space is decomposed into subspaces with a fixed number of
photons, which we will denote by $l$. The polarization
transformation in the $l$-photon subspace is given by
$(l+1)$-dimensional representation of the rotation group for spin
$l/2$. Explicitly, the decomposition of $\hat{U}({\bf \Omega})$ in
the Schwinger representation has the form:
\begin{equation}
\hat{U}({\bf \Omega}) = \bigoplus_{l=0}^{\infty}
e^{-il\alpha({\bf\Omega})} \hat{\cal
D}^{l/2}(e^{i\alpha({\bf\Omega})}{\bf \Omega})
\end{equation}
where the matrix $\hat{\cal
D}^{l/2}(e^{i\alpha({\bf\Omega})}{\bf\Omega})$ is the corresponding
element of the $(l+1)$-dimensional irreducible representation of
SU(2).

The transformation due to depolarization of the entire quantum state
$\hat{\varrho}$ of radiation contained in $N$ temporal slots is
given by the integral:
\begin{equation}
\hat{\varrho} \mapsto \int_{\text{U(2)}} d{\bf \Omega} [\hat{U}({\bf
\Omega})]^{\otimes N}\hat{\varrho} [\hat{U}^\dagger({\bf
\Omega})]^{\otimes N} \label{Eq:collno}
\end{equation}
where $d{\bf \Omega}$ is the invariant Haar measure in U(2). We are
now interested in identifying the subspaces of the entire Hilbert
space that preserve quantum coherence. For this purpose, we need to
expand the $N$-fold tensor product of the unitary transformation
$\hat{U}({\bf \Omega})$ into a direct sum of irreducible
representations. If we first divide the $N$-slot Hilbert space into
sectors containing in total exactly $L$ photons, this expansion will
in general take the following form:
\begin{equation}
\label{Eq:UNL} [\hat{U}({\bf \Omega})]^{\otimes N} =
\bigoplus_{L=0}^{\infty} e^{-iL\alpha({\bf\Omega})} \bigoplus_{j=(L
\bmod 2)/2}^{L/2} K^{j}_{NL} \cdot \hat{\cal
D}^{j}(e^{i\alpha({\bf\Omega})}{\bf \Omega}).
\end{equation}
In the above expression, the integers $K^{j}_{NL}$ define how many
times the spin-$j$ representation appears in the sector of $L$
photons distributed between $N$ slots, and we have symbolically
denoted by a dot the operation of taking a direct sum of
$K^{j}_{NL}$ replicas of this representation.

If we now insert the decomposition given in Eq.~(\ref{Eq:UNL}) into
Eq.~(\ref{Eq:collno}), the invariant integration over ${\bf\Omega}$
will completely remove coherence between subspaces with different
$L$, and also between any two sectors with different $j$. The latter
fact is a result of the canonical property of the products of
rotation matrix elements integrated over SU(2). Consequently, the
sector of the Hilbert space with given $L$ and $j$ can be treated
from the point of view of quantum information processing
applications as a noiseless subsystem with dimensionality
$K^{j}_{NL}$. This system can be used for example to implement
quantum cryptography protocols based on qudits \cite{qudits}, or any
other quantum communication task. Another application for which the
above decomposition is relevant is the transmission of classical
information \cite{BartRudoPRL03,BallDragPRA04}. Then the system
comprising $N$ slots and containing at most $L$ photons can be used
to encode $\sum_{L'=0}^{L} \sum_{j=(L'\bmod 2)/2}^{L'/2}
K_{NL'}^{j}$ distinguishable classical messages.

Let us now find and discuss explicit values of multiplicities
$K^{j}_{NL}$. First, we will analyze a restricted case when at most
only one photon is allowed to occupy a temporal slot. We can then
use the well-known result that for an ensemble of $L$ qubits the
allowed values of $j$ are $(L \bmod 2)/2, \ldots, L/2$, and that the
spin-$j$ algebra appears $\frac{2j+1}{L+1}{ L + 1 \choose L/2 - j}$
times in a direct-sum decomposition \cite{Nspinshalf}. When the
qubits are realized as polarization states of single photons and we
have additional freedom to distribute the photons between $N$
temporal slots, these values are multiplied by the factor ${N
\choose L}$ which gives the number of configurations to occupy $L$
out of $N$ slots. Thus the dimensionality $\tilde{K}^{j}_{NL}$ of a
spin-$j$ noiseless subsystem (labelled with a tilde to distinguish
it from the general scenario with no constraints on the number of
photons in a slot) is given by
\begin{equation}
\tilde{K}^{j}_{NL} = {N \choose L} \frac{2j+1}{L+1} { L + 1 \choose
L/2 - j}.
\end{equation}
The dimensionalities of noiseless subsystems for $N,L \le 6$ are
collected in Table~\ref{Tab:OnePhoton}. It is instructive to compare
these numbers with the case of pure phase encoding, when all the
input photons are prepared in an identical polarization. This
implies that their joint state of polarization belongs to the
completely symmetric subspace with the highest value of $j$, equal
to $ L/2$. Therefore phase-encoding schemes are included in
Table~\ref{Tab:OnePhoton} as entries with $j=L/2$, and  for clarity
they have been underlined. An interesting question is whether the
dimension of these subspaces can be enhanced by preparing input
photons in non-trivial polarization states. The answer is positive
for $N \ge 4$ and given by the highest entry in each column marked
with an asterisk. These entries correspond to hybrid encodings,
where both the phase and the polarization are exploited to protect
quantum coherence. The first non-trivial hybrid noiseless subsystem
occurs for three photons in four temporal slots, and it is obtained
by extending the $j=1/2$ subspace of three qubits by an additional
empty temporal slot. In general, a simple analysis of the sign of
the difference $\tilde{K}^{j}_{NL} - \tilde{K}^{j-1}_{NL}$ shows
that for given $N$ and $L$ the largest subsystem is obtained for the
highest integer ($L$ even) or half-integer ($L$ odd) value of $j$
satisfying $j \le \sqrt{L+2}/2$. Therefore, the largest noiseless
subsystem will usually have a hybrid character.

\begin{table}
\caption{Multiplicities $\tilde{K}^j_{NL}$ of spin-$j$
representations for $L$ photons distributed between $N$ slots with
at most one photon occupying each slot. Asterisked entries denote
hybrid noiseless subsystems with the highest dimensionality in each
column.}\label{Tab:OnePhoton}
\begin{ruledtabular}
\begin{tabular}{rllrrrrrr}
& $L$ & spin & \multicolumn{5}{c}{number of slots $N$} \\
\cline{4-9} & \hspace*{10mm} &
\hspace*{2mm}\phantom{$j=5/2$}\hspace*{2mm} & \makebox[4mm][r]{2} &
\makebox[7mm][r]{3} & \makebox[7mm][r]{4} & \makebox[7mm][r]{5} &
\makebox[7mm][r]{6} & \hspace*{1mm} \\\hline & 1
&$j=1/2$&$\underline{2}$&$\underline{3}$&$\underline{4}$&$\underline{5}
$&$\underline{6}$ & \\\hline &
2&$j=1$&$\underline{1}$&$\underline{3}$&$\underline{6}$&$\underline{10}
$&$\underline{15}$ & \\
& 2 & $j=0$ &1&3&6&10&15& \\
\hline & $3$ & $j=3/2$ & &
$\underline{1}$&$\underline{4}$&$\underline{10}$&$\underline{20}$ & \\
& $3$ & $j=1/2$ & & $2$ & $8\makebox[0pt][l]{$^{\star}$}$
&$20\makebox[0pt][l]{$^{\star}$}$&40 & \\
\hline & 4 & $j=2$ & & & $\underline{1}
$&$\underline{5}$&$\underline{15}$ & \\
& 4 & $j=1$&&&3&15&$45\makebox[0pt][l]{$^{\star}$}$ & \\
& 4 & $j=0$   & & & 2 & 10 & 30 & \\
\hline
& 5 & $j=5/2$ & & & & $\underline{1}$ & $\underline{6}$ & \\
& 5 & $j=3/2$ &&&&4&24 & \\
& 5 & $j=1/2$ &&&&5&30 & \\
\hline
& 6 & $j=3$   & & & & & $\underline{1}$ & \\
& 6 & $j=2$   & & & & & 5 & \\
& 6 & $j=1$   & & & & & 9 & \\
& 6 & $j=0$   & & & & & 5 &
\end{tabular}
\end{ruledtabular}
\end{table}

It is instructive to identify some earlier proposals in the
literature with entries in Table~\ref{Tab:OnePhoton}. For example,
the two protocols for quantum key distribution proposed in
Ref.~\cite{BoilGottPRL04} are based on the cases $N=L=4$ and $N=L=3$
with the corresponding values of $j$ equal respectively to $0$ and
$1/2$, and two-dimensional noiseless subsystems. The essential
advantage of these protocols is that they do not require a phase
reference, in contrast to the most straightforward phase encoding
scheme $N=2,L=1$ that gives a noiseless subsystem of the same
dimension. Similarly, the alignment-free test of Bell's inequalities
proposed by Cabello \cite{CabePRL03} uses $N=L=4$, and
Table~\ref{Tab:OnePhoton} suggests that a similar scheme should be
possible also for $N=L=3$.

\begin{figure}
\epsfig{file=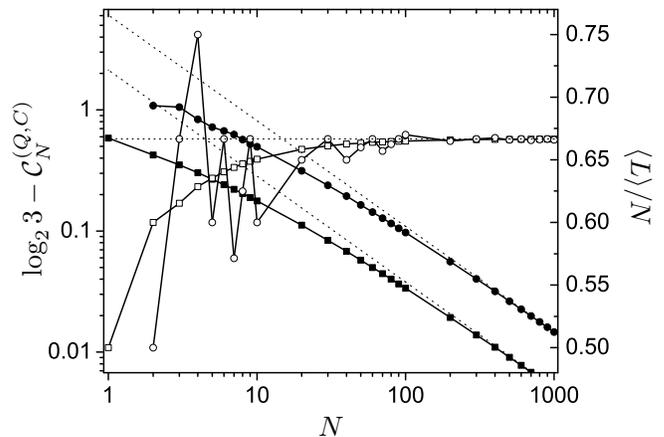,width=3.375in} \caption{Asymptotics of the
capacities ${\cal C}^{(Q)}_N$ ($\bullet$) and ${\cal C}^{(C)}_N$
({\tiny $\blacksquare$}) with the corresponding average numbers of
photons $\langle L \rangle /N$ for the case of quantum ($\circ$) and
classical ({\tiny $\square$}) communication.}
\label{Fig:Asymptotics}
\end{figure}

Given present advances in the area of scalable linear-optics quantum
computing it is interesting to investigate the asymptotics of hybrid
noiseless subsystems for large $N$. Two relevant figures of merit
are the average number of logical qubits per slot that can be
encoded in a sequence of $N$ temporal slots, given by the
expression:
\begin{equation}
{\cal C}^{(Q)}_N = \frac{1}{N} \log_{2} \left( \max_{L,j}
\tilde{K}^{j}_{NL} \right)
\end{equation}
and the classical communication capacity of a train of $N$ slots
renormalized by the number of slots:
\begin{equation}
{\cal C}^{(C)}_N = \frac{1}{N} \log_2 \left( \sum_{L=0}^{N}
\sum_{j=(L\bmod 2)/2}^{L/2} \tilde{K}_{NL}^{j} \right).
\end{equation}
We found numerically by calculating these two quantities up to
$N=10^{4}$ that they tend to the same constant value $\log_2 3 $,
with asymptotics following a power law with the approximate value of
the exponent ${\cal C}^{(Q,C)}_N \approx \log_2 3 - {\cal
O}(N^{-0.84})$, as shown in Fig.~\ref{Fig:Asymptotics}. In the same
graph, we also depict the average number of photons per slot that
are required to achieve these capacities, with the same asymptotic
limit of $\langle L \rangle /N =2/3$. Thus we see that
asymptotically the system behaves as if composed of a train of $N$
noiseless qutrits, each spanned by the vacuum state and two
one-photon states with orthogonal polarizations.

Let us now relax the constraint of having at most one photon per
temporal slot and analyze the general scenario in which photons can
be arbitrarily allocated in the temporal slots. We will derive a
closed recursion formula for the multiplicities $K_{NL}^{j}$. First,
let us simplify the notation by rewriting Eq.~(\ref{Eq:UNL})
symbolically in terms of the corresponding algebras of the noise
operators:
\begin{equation}
\label{Eq:ANL} {\cal A}_{NL}=\bigoplus_{j=(L \bmod 2)/2}^{L/2}
K^{j}_{NL} \cdot {\cal A}^{j}
\end{equation}
Here ${\cal A}_{NL}$ is the algebra for $L$ photons in $N$ slots,
and the ${\cal A}^{j}$ denote standard irreducible spin-$j$
algebras. The recursion formula will be based on a relation linking
the representation multiplicities $K^{j}_{NL}$ for $L$ photons
distributed between $N$ slots to those describing the system with
the number of slots reduced by one. When we distribute $L$ photons
between $N$ slots, we can put an arbitrary number $L' \le L$ of them
into $N-1$ slots we had before, and the remaining $L-L'$ photons
into the additional slot. Obviously, the polarization transformation
of these $L-L'$ photons in the new slot is governed by the algebra
${\cal A}^{(L-L')/2}$. Assuming that the decomposition of the
algebra ${\cal A}_{N-1,L'}$ for the photons in the remaining $N-1$
slots is known in the form of Eq.~(\ref{Eq:ANL}), this gives us the
following relation:
\begin{equation}
{\cal A}_{NL}= \bigoplus_{L'=0}^{L} {\cal A}^{(L-L')/2} \otimes
\left( \bigoplus_{j'= (L' \bmod 2)/2}^{L'/2} K^{j'}_{N-1,L'} \cdot
{\cal A}^{j'}\right).
\end{equation}
We can now use the standard decomposition of the tensor product of
spin algebras:
\begin{equation}
{\cal A}^{(L-L')/2} \otimes {\cal A}^{j'} = \bigoplus_{j=\left|
(L-L')/2 - j' \right|}^{(L-L')/2 + j'} {\cal A}^j
\end{equation}
to obtain an expression for ${\cal A}_{NL}$ in the form of a triple
direct sum:
\begin{equation}
\label{Eq:L'j'} {\cal A}_{NL}= \bigoplus_{L'=0}^{L} \bigoplus_{j'=
(L' \bmod 2)/2}^{L'/2} \bigoplus_{j=| (L-L')/2 - j' |}^{(L-L')/2 +
j'} K^{j'}_{N-1,L'} {\cal A}^j
\end{equation}
We now need to rearrange the summations in order to bring the above
expression to the form of Eq.~(\ref{Eq:ANL}). This can be done by
replacing the first two summation indices $L'$ and $j'$ by a pair of
new integer parameters $\mu=L'/2+j'$ and $\nu=L'/2-j'$. Then the
two-dimensional grid $(L',j')$ of the summation points with
$L'=0,1,\ldots, L$ and $j'=(L' \bmod 2)/2, (L' \bmod 2)/2+1, \ldots,
L'/2$ can be written as $(\mu+\nu,\frac{\mu-\nu}{2})$ with the new
parameters running through ranges $\nu=0,1,\ldots, \lfloor L/2
\rfloor$ and $\mu = \nu, \nu+1, \ldots, L-\nu$. In the new
parametrization, Eq.~(\ref{Eq:L'j'}) takes the form:
\begin{equation}
\label{Eq:munu} {\cal A}_{NL}= \bigoplus_{\nu=0}^{\lfloor L/2
\rfloor} \bigoplus_{\mu=\nu}^{L-\nu}
\bigoplus_{j=|L/2-\mu|}^{L/2-\nu} K^{(\mu-\nu)/2}_{N-1,\mu+\nu}
{\cal A}^j.
\end{equation}
If we now want to perform first the summation over $j$, we need to
identify the allowed ranges of $\mu$ and $\nu$ as a function of $j$.
Considering the limits of the sum over $j$ in Eq.~(\ref{Eq:munu}),
we obtain constraints on $\mu$ and $\nu$ in the form  $\nu \le
L/2-j$ and $ L/2-j \le \mu \le L/2+j$. These constraints are always
stronger than the original limits of the sums over $\mu$ and $\nu$
in Eq.~(\ref{Eq:munu}), and they can replace the latter.
Consequently, swapping the order of summations gives the following
recursion formula for the multiplicities $K^{j}_{NL}$:
\begin{equation}
K^{j}_{NL} = \sum_{\nu=0}^{L/2 -j}  \sum_{\mu=L/2 -j}^{L/2+j}
K^{(\mu-\nu)/2}_{N-1,\mu+\nu}.
\end{equation}
In Table~\ref{Tab:ManyPhotons} we calculate explicitly the values of
$K^{j}_{NL}$ for $N \le 8 $ and $L \le 4$. As before, for each pair
of $N$ and $L$ the underlined entry with the highest $j=L/2$
corresponds to pure phase encoding which can be used as a benchmark
to indicate whether exploiting the polarization degree of freedom
can yield a larger noiseless subsystem. The optimal noiseless
subsystems for given $N$ and $L$ in most cases turn out to be
hybrid, as shown by asterisked entries in
Table~\ref{Tab:ManyPhotons}.

\begin{table}
\caption{Multiplicities $K^j_{NL}$ for $L$ photons distributed over
$N$ slots, including arbitrary multiphoton states in each temporal
slot. Underlined entries correspond to pure phase-encoding schemes,
whereas asterisks mark optimal hybrid noiseless subsystems for given
$N$ and $L$.}\label{Tab:ManyPhotons}
\begin{ruledtabular}
\begin{tabular}{rllrrrrrrrr}
& $L$ & spin & \multicolumn{8}{c}{number of slots $N$} \\
\cline{4-10} & \hspace*{8mm} &
\hspace*{2mm}\phantom{$j=5/2$}\hspace*{2mm} & \makebox[4mm][r]{2} &
\makebox[7mm][r]{3} & \makebox[7mm][r]{4} & \makebox[7mm][r]{5} &
\makebox[7mm][r]{6} & \makebox[7mm][r]{7} & \makebox[7mm][r]{8}
&\hspace*{1mm}
\\\hline &
2&$j=1$&$\underline{3}$&$\underline{6}$&$\underline{10}$&$\underline{15}
$&$\underline{21}$ & $\underline{28}$ & $\underline{36}$ &\\
& 2 & $j=0$ &1&3&6&10&15&21&28& \\
\hline & $3$ & $j=3/2$ &\underline{4} &
$\underline{10}$&$\underline{20}$&$\underline{35}$&$\underline{56}$
&
$\underline{84}$ & $\underline{120}$ & \\
& $3$ & $j=1/2$ & $2$ & $8$ & $20$ &
$40\makebox[0pt][l]{$^{\star}$}$ & $70\makebox[0pt][l]{$^{\star}$}$
& $112\makebox[0pt][l]{$^{\star}$}$ &
$168\makebox[0pt][l]{$^{\star}$}$ & \\
\hline & 4 & $j=2$ & $\underline{5}$ & $\underline{15}$ &
$\underline{35} $&$\underline{70}$&$\underline{126}$ &
$\underline{210}$ & $\underline{330}$
&\\
& 4 & $j=1$&$3$&$15$&$45\makebox[0pt][l]{$^{\star}$}$&
$105\makebox[0pt][l]{$^{\star}$}$&$210\makebox[0pt][l]{$^{\star}$}$ & $378\makebox[0pt][l]{$^{\star}$}$ & $630\makebox[0pt][l]{$^{\star}$}$ &\\
& 4 & $j=0$   & $1$&$6$ & $20$ & $50$ & $105$ & $196$ & $336$ &
\end{tabular}
\end{ruledtabular}
\end{table}

In conclusion, we have analyzed the structure of the Hilbert space
for optical radiation contained in an arbitrary number of temporal
modes and undergoing a collective depolarization, which is one of
the dominant decoherence mechanisms in optical fibers. We have
demonstrated that optimizing communication capacity requires in
general a combination of phase and polarization encodings into joint
hybrid schemes. Although practical implementation of these schemes
may require complex preparation and measurement procedures, present
experimental effort towards the realization of scalable
linear-optics quantum computing allows for optimism that the
necessary techniques will become available. As the hybrid noiseless
subsystems utilize relative phases between pulses in a temporal
sequence, the communicating parties require in general a shared
phase reference. An interesting question is whether for hybrid
subsystems this requirement can be circumvented in a way similar to
the autocompensation technique introduced by Walton {\em et al.}
\cite{OWA}.

We acknowledge useful discussions with J.-C. Boileau, R. Laflamme,
and I. A. Walmsley. This research was supported by the UK
Engineering and Physical Sciences Research Council and by Polish
Committee for Scientific Research, Project No. PBZ KBN 043/P03/2001.


\begin{thebibliography}{99}

\bibitem{GisiRiboRMP02}
N. Gisin, G. Ribordy, W. Tittel, and H. Zbinden, Rev. Mod. Phys.
{\bf 74}, 145 (2002).

\bibitem{BouwPanNAT97}
D. Bouwmeester, J. W. Pan, K. Mattle, M. Eibl, H. Weinfurter, and A.
Zeilinger, Nature {\bf 390}, 575 (1997).

\bibitem{TimeBin}
J. Brendel, N. Gisin, W. Tittel, and H. Zbinden, Phys. Rev. Lett.
{\bf 82}, 2594 (1999); W. Tittel, J. Brendel, H. Zbinden, and N.
Gisin, {\em ibid.} {\bf 84}, 4737 (2000) W. Tittel, H. Zbinden, and
N. Gisin, Phys. Rev. A {\bf 63}, 042301 (2001).

\bibitem{HughMorgJMO00}
R. J. Hughes, G. L. Morgan, and C. G. Peterson, J. Mod. Opt. {\bf
47}, 533 (2000).

\bibitem{PlugandPlay}
A. Muller, T. Herzog, B. Huttner, W. Tittel, H. Zbinden, and N.
Gisin, Appl. Phys. Lett. {\bf 70}, 793 (1997); D. S. Bethune and W.
P. Risk, IEEE J. Quantum Electron. {\bf 36}, 240 (2000).

\bibitem{OWA}
Z. D. Walton, A. F. Abouraddy, A. V. Sergienko, B. E. A. Saleh, and
M. C. Teich, Phys. Rev. A {\bf 67}, 062309 (2003); Phys. Rev. Lett
{\bf 91}, 087901 (2003).

\bibitem{Noiseless}
For a review, see D. A. Lidar and K. B. Whaley, in {\it Irreversible
Quantum Dynamics,} edited by F. Benatti and R. Floreanini (Springer
Lecture Notes in Physics vol. 622, Berlin, 2003), p. 83.


\bibitem{BoilGottPRL04}
J.-C. Boileau, D. Gottesman, R. Laflamme, D. Poulin, and R. W.
Spekkens, Phys. Rev. Lett. {\bf 92}, 017901 (2004).

\bibitem{CabePRL03}
A. Cabello, Phys. Rev. Lett. {\bf 91}, 230403 (2003).

\bibitem{BourEiblPRL04}
M. Bourennane, M. Eibl, S. Gaertner, C. Kurtsiefer, A. Cabello, and
H. Weinfurter, Phys. Rev. Lett. {\bf 92}, 107901 (2004).

\bibitem{BartRudoPRL03}
S. D. Bartlett, T. Rudolph, and R. W. Spekkens, Phys. Rev. Lett.
{\bf 91}, 027901 (2003).

\bibitem{SchwingerRepresentation}
J. Schwinger, reprinted in {\em Quantum Theory of Angular Momentum},
edited by L. C. Biederharn and H. van Dam (Academic, New York,
1965).


\bibitem{qudits}
N.J. Cerf, M. Bourennane, A. Karlsson, N. Gisin, Phys. Rev. Lett.
{\bf 88}, 127902 (2002); A. Ac\'{\i}n, N. Gisin, V. Scarani, Quant.
Inf. Comp. {\bf 3}, 563 (2003).

\bibitem{BallDragPRA04}
J. Ball, A. Dragan, and K. Banaszek, Phys. Rev. A {\bf 69}, 042324
(2004); K. Banaszek, A. Dragan, W. Wasilewski, and C. Radzewicz,
Phys. Rev. Lett. {\bf 92}, 257901 (2004).


\bibitem{Nspinshalf}
R. H. Dicke, Phys. Rev. {\bf 93}, 99 (1954).

\end{thebibliography}
\end{document}